\documentclass[11pt]{JHEP3}
\usepackage{amssymb}
\usepackage{bbold}
\usepackage{graphicx}

\let\oldsqrt\sqrt
\def\sqrt{\mathpalette\DHLhksqrt}
\def\DHLhksqrt#1#2{%
\setbox0=\hbox{$#1\oldsqrt{#2\,}$}\dimen0=\ht0
\advance\dimen0-0.2\ht0
\setbox2=\hbox{\vrule height\ht0 depth -\dimen0}%
{\box0\lower0.4pt\box2}}
\def\Dslash{\,\,{\raise.15ex\hbox{/}\mkern-12mu D}}
\def\Dbarslash{\,\,{\raise.15ex\hbox{/}\mkern-12mu {\bar D}}}
\def\delslash{\,\,{\raise.15ex\hbox{/}\mkern-9mu \partial}}
\def\delbarslash{\,\,{\raise.15ex\hbox{/}\mkern-9mu {\bar\partial}}}
\def\pslash{\,\,{\raise.15ex\hbox{/}\mkern-11mu p}}
\def\qslash{\,\,{\raise.15ex\hbox{/}\mkern-9mu q}}
\def\calDslash{\,\,{\raise.15ex\hbox{/}\mkern-12mu {\cal D}}}

\usepackage{amsmath}
\usepackage{amssymb}
\usepackage{mathrsfs}
\usepackage{amsfonts}

\title{Enhanced (p)reheating in DBI Inflation}
\author{A. C. Davis\\ E-mail:  \email{A.C.Davis@damtp.cam.ac.uk}\\Department of Applied Mathematics and Theoretical Physics\\
Centre for Mathematical Sciences\\
Cambridge CB2 0WA, United Kingdom} 
\author{Raquel H. Ribeiro\\ E-mail:  \email{R.Ribeiro@damtp.cam.ac.uk}\\ Department of Applied Mathematics and Theoretical Physics\\
Centre for Mathematical Sciences\\
Cambridge CB2 0WA, United Kingdom}

\abstract{
We study preheating in DBI hybrid inflation. Preheating occurs since the matter fields are non-minimally coupled to the inflaton. Despite the coupling being small, preheating happens as a narrow parametric resonance in which matter fields and inflaton modes are exponentially created. We show that, within the approximations used, the process is very efficient at late times. The resonance stage gives way to the decay of the inflaton into bosonic or fermionic degrees of freedom and the corresponding reheating temperature is determined.}

\begin{document}
\section{Introduction}
Recent developments in string theory have resulted in attempts to construct string inspired models of inflation. Whilst progress has been made in brane inflation models \cite{Dvali:1998pa, Cline:2006hu, Burgess:2006uv, Baumann:2009ni}, the theory of reheating at the end of inflation 
in such models is still in its infancy \cite{Shtanov:1994ce, Kofman:1997yn, Barnaby:2004gg, Chialva:2005zy, Langfelder:2006vd}. Most string inspired models of inflation have concentrated on the form
of the potential. However, in an interesting paper \cite{Silverstein:2003hf} Silverstein and Tong proposed a novel D-cceleration setup within the context of Brane Inflation theory. 
This opened the possibility for fast-roll inflation to take place, circumventing the fine-tuning problems. Their model uses non-standard kinetic terms for the inflaton field of the Dirac-Born-Infeld (\textit{aka} DBI) type, motivated since the DBI action provides the low-energy description of a  
D3-brane in an AdS space \cite{Aharony:1999ti}.

A novel feature of DBI inflation is that it predicts non-Gaussian primordial spectrum \cite{Chen:2005fe} for the scalar perturbations which can then be confronted with observational data. In addition, the speed of sound is time dependent \cite{Alishahiha:2004eh}.
Available data suggests tighter constraints on the parameters of the brane inspired inflationary 
models. This opens up the exciting possibility for the Planck satellite \cite{Chen:2006nt} \cite{Peiris:2007gz} to shed some light on other models besides the standard single-field inflationary ones \cite{Seery:2005wm}. Whereas the inflationary perturbations have been studied in DBI inflation, up until now there has been no attempt to study the theory of reheating. In this paper we study a simple model of reheating within DBI inflation. 

Reheating occurs at the end of inflation when the energy stored in the inflaton field is transferred to matter and radiation due to the oscillation of the inflaton around the minimum of the potential. Due to the coupling with matter fields this oscillation can lead to parametric resonance, resulting in the explosive non-perturbative growth of particles. When the final state is thermalized, it is said that reheating has occurred. 
This preheating phase has been studied extensively for standard models of inflation \cite{Traschen:1990sw, Shtanov:1994ce, Kofman:1997yn, Felder:1998vq}. The extension to models with non-standard kinetic terms is in its early stages, though non-standard matter fields have been studied recently \cite{Lachapelle:2008sy}. In this paper we take a different approach. We consider DBI inflation with the inflaton field coupled to matter fields. As a first step we take the matter fields to have standard kinetic terms. 

Our model is a variant of hybrid inflation \cite{Linde:1993cn} and is motivated by $D3/D7$ brane inflation. We show the modifications to the preheating effect that need to be taken into account, when the inflaton field has DBI action. The theory is described by

\begin{displaymath}
 S=\int{\sqrt{-g_m} \left(\, \mathcal{L}_0 \ +\  \dots     \right)  } \ \ ,
\end{displaymath} 
with signature $(-,+,+,+)$ for the metric and
\begin{equation}
\mathcal{L}_0 \, = \, -\, \left[\dfrac{1}{f(\phi)}\sqrt{1+f(\phi)\, g^{\mu\nu}\partial_{\mu}\phi \partial_{\nu}\phi}- \dfrac{1}{f(\phi)}  \right]-V(\phi,\chi)-\dfrac{1}{2}\partial_{\mu}\chi \partial^{\mu} \chi \ ,
\label{eq:lagrangian}
\end{equation} 
where the inflaton field $\phi$ is measured in units of the string coupling, $g_s$.  $f(\phi)$ is the warp factor, which is inversely proportional to the brane tension and $\chi$ is the matter field. 
The potential used is
\begin{equation}
V\, \equiv \, V\left(\phi, \chi\right)=2g^2 \phi^2 \chi^2 +\dfrac{g^2}{2}\left(\chi^2-\zeta_3\right)^2 \ \ ,
\label{eq:potential}
\end{equation} 
where $\zeta_3$ is the symmetry breaking term. Any one-loop corrections are sub-dominant and will not be considered.

The modifications due to the DBI kinetic term will be studied and compared to the case with standard kinetic terms and the same potential \cite{Brandenberger:2008if}. This way we gain insight into how (p)reheating is modified due to non-standard DBI kinetic terms. This paper is organised as follows: first we study the cosmological evolution of the inflaton field as it rolls down the potential from the false to the true vacuum configuration. We then assess the differences in the (p)reheating process, analysing the possibility of broad/narrow resonance occurring. We compare the efficiency with respect to the usual case. Finally, we discuss our results and we summarise the underlying implications. 

In what follows we will use natural Planckian units, in which the reduced Planck mass is $\mu_{Pl}:=\dfrac{M_{Pl}}{\sqrt{8\pi}}\simeq 2.4 \times 10^{18}GeV$.

\section{Dynamics of Inflation}
In this section, we compute the cosmological evolution of the scalar fields and investigate the circumstances under
which (p)reheating occurs.

\subsection{Equations of Motion}
The equations of motion for both fields are 
\begin{equation}
\ddot{\phi}\ +\ 3\, H  \, \dfrac{\dot{\phi}}{\gamma^2} \ -\ \dfrac{f'}{f^2} \ + \ \dfrac{3}{2}\, \dfrac{f'}{f}\, \dot{\phi}^2 \ + \ \dfrac{1}{\gamma^3}\, \left(\dfrac{f'}{f^2}+\dfrac{\partial V}{\partial \phi}  \right)\ = \ 0  \ \ ,
\label{eq:eom_inflaton}
\end{equation} 
and 
\begin{equation}
\ddot{\chi}\ +\ 3\, H  \,\dot{\chi} \ + \ \dfrac{\partial V}{\partial \chi} \ = \ 0  \ \ .
\label{eq:eom_matter}
\end{equation} 
Here
\begin{equation}
 \gamma:=\left(1-f\dot{\phi}^2  \right)^{-1/2}
\label{eq:gama}
\end{equation}
plays a similar role to that of the Lorentz contraction factor and dot denotes the derivative with respect to 
cosmological time. Current data requires $\gamma\lesssim32$ for compatibility with observation \cite{Burgess:2007pz}.

Clearly, the difficulty with these equations is that the square root from the inflaton kinetic term
makes the equations highly non-linear. To solve the equations we adopt approximation techniques. Firstly, we assume 
a cut-off throat in the infrared, so that the warp factor becomes
\begin{displaymath}
 f(\phi)=\dfrac{\lambda}{\left(\phi^2+\mu^2  \right)^2} \  \ ,
\end{displaymath}
for some new energy scale $\mu$. Since we are interested in the late time behaviour towards the very end of inflation, 
then $\phi \to 0$, which allows us to simplify the warp factor as
\begin{displaymath}
  f(\phi)=\dfrac{\lambda}{\mu^4} \  \ .
\end{displaymath}
This assumption will be in agreement with the further approximations in the next sections.

The modification of the equations of motion for the inflaton field result in a modification to the the usual 
preheating mechanism. In particular, the solutions to the equations of motion of the inflaton field acquire a 
correction to the canonical harmonic oscillator equation. We make this explicit in the next section, where such 
(perturbative) correction is computed.

\subsection{Reaching the bottom of the potential}
At the start of inflation, $\left< \chi \right>=0$ and the inflaton 
starts rolling down the potential. As $\phi$ goes down the potential, slow-roll conditions are no longer valid 
(since the field speeds up when meeting a steeper region of the potential) and it approaches its bottom, 
overshooting and it starts oscillating. Eventually, both fields should assume their vacuum expectation values, namely
\begin{displaymath}
 \left<\phi\right>=0 \ \ \ \textrm{and} \ \ \ \left<\chi\right>=\sqrt{\zeta_3} \ \ ,
\end{displaymath} 
for which they minimise the potential and violate slow-roll conditions.

Next, we will study the fluctuations of the fields around their background solutions.

\section{Preheating}\label{Preheating}
In our study we neglect the Hubble friction term since the expansion time is much larger than the one we are interested 
in. We perform a Taylor expansion of the $\gamma$ factor, keeping first order terms, and thus neglect 
$\mathcal{O}\left(f^2\, \dot{\phi}^4\right)$ since this is higher order. This is because the presence of the branch 
cut in the denominator of (\ref{eq:gama}) demands a limiting speed of the D3-brane moving along the throat to be 
\begin{equation}
 \left| \, \dot{\phi} \, \right| \leq \dfrac{\mu^2}{\sqrt{\lambda}} \ \ ,
\label{eq:speed_limit}
\end{equation}
for $\phi\lesssim \mu$. The modulus sign accounts for the fact that $\phi$ decreases in time as it gets closer to the 
tip of the throat. Hence, the contributions coming from these higher order terms are negligible.
Then, (\ref{eq:eom_inflaton}) becomes manifestly simplified and 
\begin{equation}
 \ddot{\phi}+\left[1-\dfrac{3}{2}\lambda\dfrac{\dot{\phi}^2}{\mu^4}\right]  \, 4g^2\zeta_3 \phi \simeq0  \ \ .
\label{eq:simplified_eom_inflaton}
\end{equation} 
Considering the extra term to be a small perturbation to the homogeneous harmonic oscillator equation, we require 
$f\, \dot{\phi}^2<2/3$, which is a stronger requirement but still compatible with (\ref{eq:speed_limit}). 
We hence get a correction to the usual preheating equation of motion for the inflaton, with solution given by
\begin{displaymath}
 \phi= \mathcal{A} \, \cos{\left(\omega \, t  \right)} \ \ ,
\end{displaymath}
with $\mathcal{A}$ and $\omega=2g\sqrt{\zeta_3}$ being the oscillation amplitude and frequency, respectively. It follows that the solution to (\ref{eq:simplified_eom_inflaton}) may then be obtained by assuming an additional perturbative contribution to the latter equation, yielding
\begin{equation}
  \phi\simeq \mathcal{A}\cos\left(\omega t\right) +\dfrac{3}{4}\dfrac{\lambda}{\mu^4}\mathcal{A}^3 \omega^2 \sin^2\left(\omega t\right) \cos\left(\omega t\right) \ \ ,
\label{eq:correction_inflaton}
\end{equation} 
up to a normalisation factor. This \textit{ansatz} is still a periodic function of time, so that one can trace its time evolution throughout an unlimited number of periods. The strength of this perturbation is parametrised by the coefficient sitting outside the periodic function and we note that, due to the Hubble friction term, the amplitude of the oscillations is expected to be damped. Thus, we expect $\dfrac{3}{4}\dfrac{\lambda}{\mu^4}\mathcal{A}^3 \omega^2<\mathcal{A}$, which is in agreement with the condition for the stronger restriction $f\, \dot{\phi}^2<2/3$.

Slow-roll conditions stop being valid when, for small values of the matter field $\chi$, $\phi<\sqrt{\dfrac{\zeta_3}{2}}$ and a tachyonic instability sets in. From here and to lowest order in the approximation scheme, the oscillatory background solution for the matter field (\ref{eq:eom_matter}) will follow a linear trajectory according to 
\begin{equation}
 \chi=\sqrt{\zeta_3}+\bar{\chi} \ \ \ \textrm{with} \ \ \ \bar{\chi}=-\sqrt{2}\, \phi \ \ , 
\label{eq:trajectory}
\end{equation}
which can be shown to obey the equations of motion (\ref{eq:eom_inflaton}) and (\ref{eq:eom_matter}). This suggests the correct rescaling in (\ref{eq:correction_inflaton}) for the inflaton background solution. Thus, we consider the (normalised) classical background solutions to the equations of motion (\ref{eq:eom_inflaton}) and (\ref{eq:eom_matter}) to be
\begin{equation}
\left\{
\begin{array}{ll}
\sqrt{2}\,\tilde{\phi}&=\mathcal{A}\cos\left(\omega t\right) +3\,g^2\,\dfrac{\lambda}{\mu^4}\mathcal{A}^3 \zeta_3 \sin^2\left(\omega t\right) \cos\left(\omega t\right) \\ \\
\ \ \ \tilde{\chi}&=\sqrt{\zeta_3}- \mathcal{A}\cos\left(\omega t\right) -3\,g^2\,\dfrac{\lambda}{\mu^4}\mathcal{A}^3 \zeta_3 \sin^2\left(\omega t\right) \cos\left(\omega t\right) \ \ .
\end{array}
\right. \ \ 
\label{eq:background_solutions}
\end{equation}

\subsection{Fluctuations around the background solutions}
We now consider the approximate classical fluctuations in the direction of the background solutions in (\ref{eq:background_solutions}), defined by
\begin{equation}
\left\{
\begin{array}{ll}
\delta\phi:=\phi-\tilde{\phi} \\ \\
\delta\chi:=\chi-\tilde{\chi} \ \ .
\end{array}
\right. \ \ 
\label{eq:fluctuations_def}
\end{equation}
Importantly, in order to solve the modified equations of motion (\ref{eq:eom_inflaton}) and (\ref{eq:eom_matter}) we 
need to consider $\mathcal{O}\left(\mathcal{A}^3 \right)$; however, for the purpose of studying the equations of motion for the 
field fluctuations it is sufficient to consider $\mathcal{O}\left(\mathcal{A}^2\right)$ terms. Also, we consider the 
fluctuating Fourier modes in a Hubble volume with comoving number $k$ (up to a normalisation factor):
\begin{displaymath}
\left\{
\begin{array}{ll}
\phi(x)=\displaystyle{\int{d^3\, k \ \left(a_{\vec{k}}\, \phi_k(t)\, e^{-i k\ldotp x} +a^{\dagger}_{\vec{k}}\, \phi_k^{*}(t)\, e^{i k \ldotp x}   \right)}    }  \\ \\
\chi(x)=\displaystyle{\int{d^3\, k \ \left(b_{\vec{k}}\, \chi_k(t)\, e^{-i k\ldotp x} +b^{\dagger}_{\vec{k}}\, \chi_k^{*}(t)\, e^{i k \ldotp x}   \right)} } \ \ ,
\end{array}
\right. \ \ 
\end{displaymath}
where $\left\{a_{\vec{k}}, b_{\vec{k}} \right\}$ and $\left\{ a^{\dagger}_{\vec{k}}, b^{\dagger}_{\vec{k}}    \right\}$ are respectively the annihilation and creation operators, with momentum $\vec{k}$. Moreover, since we are expanding around the background solutions (\ref{eq:background_solutions}), which satisfy the equations of motion (\ref{eq:eom_inflaton}) and (\ref{eq:eom_matter}), it follows that only the quadratic terms in the potential (\ref{eq:potential}) are relevant. Hence, the fluctuations with comoving number $k$ obey the following equations of motion in the momentum space:
\begin{equation}
\left\{
\begin{array}{ll}
 \delta\ddot{\phi}_k +k^2\delta\phi_k +\left[ 1-\dfrac{3}{2}\dfrac{\lambda}{\mu^4}\dot{\tilde{\phi}}^2     \right] \, \left[8g^2\, \tilde{\phi}\, \tilde{\chi}\, \delta\chi_k +4g^2\, \tilde{\chi}^2\delta\phi_k     \right] \simeq0 \\ \\
\delta\ddot{\chi}_k +8g^2\, \tilde{\phi}\, \tilde{\chi}\delta\phi_k+\left[k^2+2g^2\left(2\tilde{\phi}^2+3\tilde{\chi}^2-\zeta_3     \right)    \right]\delta\chi_k\simeq0 \ \ .
\end{array}
\right. \ \ 
\label{eq:fluctuations_eom1}
\end{equation}
In particular, in agreement with the linear trajectory defined in (\ref{eq:trajectory}), the fluctuations of both fields are related by
\begin{equation}
 \delta\chi_k=-\sqrt{2}\, \delta\phi_k \ \ .
\end{equation} 
Plugging it into (\ref{eq:fluctuations_eom1}) and using the \textit{ansatz} (\ref{eq:background_solutions}), we get the 
simplified equations of motion for the fluctuations of the inflaton and matter fields around their respective background 
solutions, up to the order of approximation used:
\begin{equation}
\left\{
\begin{array}{lll}
 &\delta\ddot{\phi}_k  &+\left[k^2+4g^2\left\{\zeta_3\, +\, \dfrac{3}{2} \mathcal{A}^2\left(1-\dfrac{\lambda}{\mu^4}\, g^2\zeta_3^2 \right)\right\} \right.  +\\ \\
  & &\left.-16g^2\, \sqrt{\zeta_3} \mathcal{A}\cos{\left(\omega t\right)}   +6g^2\, \mathcal{A}^2 \left( 1+\dfrac{\lambda}{\mu^4}\, g^2 \zeta_3^2     \right) \cos{\left(2\,\omega t  \right)}                            \right]    \delta\phi_k   \simeq0 \\ \\
&\delta\ddot{\chi}_k& +\left[k^2+4g^2\zeta_3+6g^2\mathcal{A}^2 \ - \ 16g^2 \, \sqrt{\zeta_3}\, \mathcal{A}\, \cos{\left(\omega t\right)} \ + \ 6g^2 \mathcal{A}^2 \cos{\left(2\omega t\right)}      \right]\, \delta \chi_k \simeq0 \ \ .
\end{array}
\right. \ \ 
\label{eq:fluctuations_eom2}
\end{equation}
%In the process of getting these equations, we have considered the fact that although the initial value for the amplitude $\mathcal{A}$ of the field oscillations starts at the order of $\sqrt{\zeta_3}$, it is clear that it will be intuitively damped by the Hubble friction term (as a long term consequence), but also by the backreaction of the matter field fluctuations which (efficiently or not) deplete energy from the inflaton field.  
We note that the equation for the fluctuations of the inflaton field resembles Hill's equation, which is usually written in the form
\begin{displaymath}
 \dfrac{d^2y}{dz^2}+\left[\theta_0+2\displaystyle{\sum_{r=1}^{+\infty}{\theta_{2r}\, \cos{\left( 2r z \right)}}}   \right] y=0 \ \ ,
\end{displaymath}
where $\theta_0$, $\theta_2$, $\theta_4$, $\dots$ are assigned parameters and $\displaystyle{\sum_{r=1}^{+\infty}{\left|\theta_{2r}  \right|}}$ converges. Indeed, introducing a rescaled dimensionless time defined as $2z:=\omega t$, it follows that the first equation in (\ref{eq:fluctuations_eom2}) may be written as 
\begin{equation}
 \delta\phi^{''}_k \ + \ \left[\theta_0+2\theta_2 \, \cos{\left(2z\right)} +2\theta_4 \, \cos{\left(4z\right)}  \right]  \delta\phi_k =0 \ \ ,
\label{eq:fluctuations_eom3}
\end{equation} 
where the prime denotes the derivative with respect to the rescaled time with the identifications
\begin{equation}
\left\{
\begin{array}{lll}
\theta_0=4+\dfrac{k^2}{g^2\zeta_3}+6\dfrac{\mathcal{A}^2}{\zeta_3}\left(1-\dfrac{\lambda}{\mu^4} g^2\zeta_3^2   \right) \\ \\
\theta_2=-8\dfrac{\mathcal{A}}{\sqrt{\zeta_3}} \ <\ \ 0 \\ \\
\theta_4=3\dfrac{\mathcal{A}^2}{\zeta_3}\, \left(1+\dfrac{\lambda}{\mu^4} g^2\zeta_3^2   \right) \ > \ 0 \ \ .
\label{eq:theta_values}
\end{array}
\right. \ \ 
\end{equation}

Hill's equation with three terms may be transformed in a type of equation which was already studied in \cite{Lachapelle:2008sy} by performing the following change of variables \cite{arscott}
\begin{equation*}
 \delta\phi_k=v_k \, e^{-\sqrt{\theta_4}}\, \cos{(2z)} \ \ ,
\end{equation*}
so that in terms of the new field, (\ref{eq:fluctuations_eom3}) is rewritten as
\begin{displaymath}
 \dfrac{d^2v}{dz^2}+4\sqrt{\theta_4}\sin{(2z)}\,\dfrac{d v}{dz}+\left[\theta_0+2\theta_4+\cos{(2z)}\left(2\theta_2+4\sqrt{\theta_4}    \right)     \right]\, v=0 \ \ .
\end{displaymath} 
This has the generic form of
\begin{equation}
\dfrac{d^2v}{dz^2}+2p\sin{(2z)}\dfrac{d v}{dz}+\left[ \mathscr{A}_k -2\bar{q} \cos{(2z)}   \right]v=0 \ \ ,
\end{equation}
with 
\begin{equation}
\left\{
\begin{array}{ll}
p&=2\sqrt{\theta_4} \\ \\
\mathscr{A}_k&=\theta_0+2\theta_4 \\ \\
\bar{q}&=-\left(\theta_2+2\sqrt{\theta_4}   \right) .
\label{eq:bridge_with_RB}
\end{array}
\right. \ \ 
\end{equation}
We observe that $\bar{q}$ can be either positive or negative, depending on the relative strength of $\theta_2$ and $\theta_4$. This equation is known in literature \cite{mclachlan:1965} and we will study this in detail in the next section.

It is intriguing that the phenomenon of inflaton particle production also takes place in preheating. This follow as a 
consequence of the DBI term. In fact, at the level of the equations of motion, (\ref{eq:simplified_eom_inflaton}) may be 
rewritten in a more suggestive way as
\begin{equation}
\ddot{\phi}+\dfrac{dV_{eff}}{d\phi}=0 \ \ ,
\end{equation} 
with $V_{eff}$ approximately given by 
\begin{equation}
 V_{eff}\simeq \dfrac{1}{\gamma^3} V
\end{equation} 
and with $V$ as defined in (\ref{eq:potential}). Therefore, the rescaled potential has now a cubic power in the field, explaining the self interaction observed here as a consequence of the DBI term. In this sense, DBI models fall in the same class as self-interaction models where resonant inflaton production is, in general, expected \cite{Greene:1997fu}.

We note that since the corrections to the inflaton fluctuations are $\mathcal{O}\left(\mathcal{A}^2\right)$
we can truncate the contributions to both the matter and inflaton equations at this order.
This amounts to neglecting the DBI correction terms to the inflaton background solution. Therefore, we would expect to 
obtain the same equations of motion for the fluctuations about the background matter field solution as in 
\cite{Brandenberger:2008if}, except that now $\mathcal{O}\left(\mathcal{A}^2\right)$ can no longer be neglected as 
being subdominant, \textit{a priori}, at this order. However, the 
matter field equations involve no corrections from DBI\footnote{this is because in the case of the fluctuations of the inflaton field the strength of the terms is not only parametrised by $\mathcal{A}$ but also by the cut-off radius $\mu$ in the term $\lambda/\mu^4$.}. This allows us to ignore these terms, obtaining the further approximate equation of motion:
\begin{equation}
 \delta\ddot{\chi}_k +\left[k^2+4g^2\zeta_3 \ - \ 16g^2 \, \sqrt{\zeta_3}\, \mathcal{A}\, \cos{\left(\omega t\right)}      \right]\, \delta \chi_k \simeq0 \ \ .
\end{equation}
This has the standard form of Mathieu's equation which is a special case of Hill's equations by taking 
$\theta_4=0$ in (\ref{eq:theta_values}). Indeed, introducing the dimensionless time scale defined via $2z=\omega t$ and denoting, like before, the derivative with respect to $z$ as a prime, then the above equation is rewritten as 
\begin{equation}
  \delta\chi^{''}_k +\left[A_k -2q \cos{(2z)}      \right]\, \delta \chi_k =0 \ \ ,
\label{eq:matter_equation}
\end{equation} 
where
\begin{equation}
\left\{
\begin{array}{ll}
A_k&=4+\dfrac{k^2}{g^2\zeta_3} \\ \\
 \ \, q&=8 \dfrac{\mathcal{A}}{\sqrt{\zeta_3}} \ \ .
\label{eq:matter_mathieu}
\end{array}
\right. \ \ 
\end{equation}
Mathieu's equation is well known (see, for example, \cite{Kofman:1997yn} and \cite{Greene:1997fu}) to be associated 
with preheating. Its features are essentially that its solutions are a linear combination of exponentially growing and decaying modes, leading to instability bands dependent on the choice of values for $A_k$ and $q$, which are in turn interpreted as a particle production phenomenon.

\subsection{Efficiency of particle production}\label{efficiency}
Although the equation for the matter field fluctuations in (\ref{eq:fluctuations_eom2}) was already obtained in \cite{Brandenberger:2008if}, the equation for the inflaton field fluctuations contains new terms, requiring a concise study 
of the orders of magnitude of the terms in (\ref{eq:fluctuations_eom3}). The main point is that the speed limit (\ref{eq:speed_limit}) for the D3-brane moving along the cut-off throat provides a natural limit for the orders of magnitude of several parameters in this equation. Together with the lowest order background solution for the inflaton field (\ref{eq:background_solutions}), it follows that
\begin{displaymath}
 0 < 2\,  \dfrac{\lambda}{\mu^4}\, g^2\, \zeta_3\, \mathcal{A}^2 \, \sin^2{(\omega t)} \lesssim \dfrac{2}{3} \ \ ,
\label{eq:criteria}
\end{displaymath} 
This, together with the decreasing amplitude of the oscillatory contributions themselves, verifies that the correction 
to the usual background solution in (\ref{eq:correction_inflaton}) is of lower order. In order to obtain an order of
magnitude we can take the average of the oscillatory solution over several periods with respect to the late time 
behaviour.  Since $\sqrt{\zeta_3}$ and $\mu$ are fixed mass scales, the order of magnitude of the term in the criteria 
will vary as $\mathcal{A}$ decreases in time. Therefore, its maximum value corresponds to the maximal oscillation 
amplitude $\mathcal{A}$ in this regime. However, at early times we cannot use the same criteria for averaging over 
periods. However, if $f\dot{\phi}^2<2/3$, then this will certainly be true when the scalar field has maximal speed, ie, at the very bottom of the potential. Furthermore, by the time the field has reached the bottom of the potential during 
the oscillation, the amplitude $\mathcal{A}$ has not changed significantly. Taking this into account, the following 
criteria should follow, for early and late times, respectively:
\begin{equation}
 0<\dfrac{\lambda}{\mu^4}g^2\, \zeta_3^2 < \dfrac{1}{3} \dfrac{\zeta_3}{\mathcal{A}^2} \ \ \textrm{and} \ \  0<\dfrac{\lambda}{\mu^4}g^2\, \zeta_3^2 < \dfrac{2}{3} \dfrac{\zeta_3}{\mathcal{A}^2} \ \ .
\end{equation} 
For now, ``late'' times means studying the oscillation over enough periods so that the averaging over the oscillatory solution is approximately $\left<\sin^2{(\omega t)}\right>\simeq 1/2$. 
%In fact, we will evaluate in section \ref{late_times} the amplitude $\mathcal{A}$ at which the late time behaviour dominates over the early time solution and then study the implications for (\ref{eq:fluctuations_eom3}).

To proceed, we need to compare the two energy scales, $\sqrt{\zeta_{3}}$ and $\mathcal{A}$. Following 
\cite{Brandenberger:2008if}, for the usual kinetic terms, the condition for efficient resonance was shown to require 
that the final amplitude $\mathcal{A}_f$ for the oscillations of the field around the minimum of the potential were 
related to the initial amplitude $\mathcal{A}_i$ by
\begin{equation}
 \dfrac{\mathcal{A}_f}{\mathcal{A}_i} \lesssim \left( \dfrac{\sqrt{\zeta_3}}{64\, \mu_{Pl}}  \right)^{1/2} \ \ .
\end{equation} 
To get an order of magnitude, we assume the limiting value for which $\dfrac{\mathcal{A}_f}{\mathcal{A}_i}\sim\dfrac{1}{40}$. This will be tested later against the condition for efficient preheating. Hence, at the very beginning of the 
oscillation, since we expect $\mathcal{A}_i\simeq\sqrt{\zeta_3}$, then
\begin{equation*}
\left. \dfrac{\zeta_3}{ \mathcal{A}^2}\right|_{i}\sim 1  \ \ ;
\end{equation*}
on the other hand, as a consequence of particle creation taking into account the contribution of the quartic 
self-interaction term in $\chi$ in draining energy from the inflaton field, we expect $\mathcal{A}$ to decrease 
significantly, so that 
\begin{equation*}
 \left. \dfrac{\zeta_3}{ \mathcal{A}^2}\right|_{f}\sim 1600 \ \ .
\end{equation*}
Thus, it follows that the term $\dfrac{\lambda}{\mu^4}g^2\, \zeta_3^2$ becomes relevant at late times, when the DBI corrections are more important.

In this spirit, we may approximate the equations of motion for the inflaton modes in (\ref{eq:fluctuations_eom2}) into two different regimes: up to a few oscillations and then the truly late time behaviour.

\subsubsection{Within a few oscillations}
At early times, the inflaton mode equation in (\ref{eq:fluctuations_eom2}) may be approximated by 
\begin{equation}
 \delta\ddot{\phi}_k  +\left[k^2+4g^2\zeta_3 +6g^2\, \mathcal{A}^2 -16g^2\, \sqrt{\zeta_3} \mathcal{A}\cos{\left(\omega t\right)}   +6g^2\, \mathcal{A}^2 \cos{\left(2\,\omega t  \right)}                            \right]    \delta\phi_k   \simeq0  \ \ ,
\label{eq:one_reduction}
\end{equation} 
for $\mathcal{A}_f \ll \mathcal{A}\lesssim \sqrt{\zeta_3}$. 

In this regime we can make further approximations by taking into consideration which terms contribute the most to this 
equation. In particular, because $\dfrac{16 g^2 \, \sqrt{\zeta_3} \, \mathcal{A}}{6g^2\, \mathcal{A}^2}\gtrsim 8/3$, the equality will only be made manifest at the beginning of the oscillation. Hence the effect of the quadratic terms in 
$\mathcal{A}$ may be neglected so that for $\mathcal{A}<\sqrt{\zeta_3}/4$, (\ref{eq:one_reduction}) takes the form
\begin{equation}
 \delta\ddot{\phi}_k  +\left[k^2+4g^2\zeta_3  -16g^2\, \sqrt{\zeta_3} \mathcal{A}\cos{\left(\omega t\right)}       \right]    \delta\phi_k   \simeq0  \ \ .
\label{eq:two_reductions}
\end{equation} 
We recover the same equation as in \cite{Brandenberger:2008if}, so we only state their main results. This is also 
equivalent to (\ref{eq:matter_equation}), so we will be seeking unstable solutions of this Mathieu's equation (this is as expected as the DBI corrections become important at later times). This solution is of narrow-resonance type,
but is efficient since the modes remain in the resonance band for sufficient time to become exponentially modified.
In fact, denoting the instability parameter with no DBI corrections by $\mu_k^0$, the result is that the unstable 
mode should grow as
\begin{equation}
 \delta\phi_k \, \sim \, e^{\frac{1}{2}\mu_k^0 \, \omega \, t} = e^{4g\mathcal{A}t} \ \ ;
\label{eq:instability_parameter_old}
\end{equation}
we identify the width of momenta of produced particles as $\triangle k =16/3 \sqrt{3}g\mathcal{A}$ with centre at $k_0=g\sqrt{3\, \zeta_3}$.

\subsubsection{At later times}\label{late_times}

After a few oscillations, we go into the late times regime for which (\ref{eq:fluctuations_eom2}) takes the form
\begin{equation}
 \delta\ddot{\phi}_k+\left( k^2+4g^2\zeta_3-6\dfrac{\lambda}{\mu^4}g^4\mathcal{A}^2\zeta_3^2+  6\dfrac{\lambda}{\mu^4}g^4\mathcal{A}^2\zeta_3^2 \, \cos{\left(2\omega t  \right)}     \right)\, \delta\phi_k\simeq \, 0 \ \ ,
\end{equation} 
which is of the form of Mathieu's equation 
\begin{equation}
 \dfrac{d^2 \delta\phi_k}{d \tilde{z}^2}+\left(\tilde{A}_k-2\tilde{q}\cos{(2\tilde{z})}   \right) \, \delta\phi_k \ = \ 0 \ \ ,
\end{equation} 
where 
\begin{equation}
\left\{
\begin{array}{ll}
 \ \, \tilde{z}&=\omega t \\ \\
\tilde{A}_k&=1+\dfrac{k^2}{4g^2\zeta_3} -\dfrac{3}{2}\, \dfrac{\lambda}{\mu^4}\, g^2\, \zeta_3\mathcal{A}^2\\ \\
 \ \, \tilde{q}&= -\dfrac{3}{4}\, \dfrac{\lambda}{\mu^4}\, g^2\, \zeta_3\mathcal{A}^2 \, < \, 0 \ \ .
\label{eq:inflaton_mathieu}
\end{array}
\right. \ \ 
\end{equation}
The criteria for narrow resonance is verified \cite{Mukhanov:2005sc}
\begin{displaymath}
 \left(\omega\right)^2  \ \gg \ g^2\, \zeta_3 - \dfrac{3}{2}\, \dfrac{\lambda}{\mu^4}\, g^4\, \zeta_3^2 \, \mathcal{A}^2 \ \geq \ \dfrac{3}{2}\, \dfrac{\lambda}{\mu^4}\, g^4\, \zeta_3^2 \, \mathcal{A}^2 \ \ .
\end{displaymath}
This essentially corresponds to $\tilde{A}_k>2 \left|\tilde{q}\right|$. The interesting behaviour of Mathieu's equation 
is due to the presence of two master frequencies: $\tilde{A}_k$ and $\left|2\tilde{q}\cos{(2\tilde{z})}\right|$. 
The interplay of these two frequencies, particularly since the second one is evolving in time,
generates interactions between the two possible regimes depending on whether $\tilde{A}_k<\left|2\tilde{q}\cos{(2\tilde{z})}\right|$ or $\tilde{A}_k>\left|2\tilde{q}\cos{(2\tilde{z})}\right|$.

The theory of narrow parametric resonance is well known (\cite{Shtanov:1994ce} and \cite{Kofman:1997yn}). In short, the resonance occurs for very narrow bands in the momentum and the most important and widest is the first one. Indeed, for such case, $\tilde{\mathcal{A}}_k \sim 1\pm\left|\tilde{q}\right|$ and so the values of $k$ which are affected by the resonance stage will be lying within the range 
\begin{displaymath}
 k \sim \omega\, \left(\dfrac{3}{2}\, \dfrac{\lambda}{\mu^4}\, g^2\, \zeta_3 \, \mathcal{A}^2  \pm \dfrac{\left|\tilde{q}\right|}{2}   \right) \ \ .
\end{displaymath}

The general solution to Mathieu's equation may be written as 
\begin{displaymath}
 \delta\phi_k(\tilde{z}) \ \sim \ a\, e^{\mu_k \tilde{z}} \, f(\tilde{z})+ b\, e^{-\mu_k \tilde{z}} \, f(-\tilde{z}) \ \ , 
\end{displaymath}
where $\mu_k$ is the instability parameter (slowly varying function of $z$) and $f(\tilde{z})$ is some periodic function in $\tilde{z}$ with period $\pi$. Solutions with $\Re(\mu)\neq 0$ are superposition of exponentially decaying and growing modes and these last ones may induce resonant behaviour. For the narrow regime, using Hill's recurrence perturbative 
method \cite{arscott},
one can show that the maximal value for $\mu_k$ is $\left. \mu_k \right|_{max}\simeq \left|\tilde{q}\right|/2$, ie, 
\begin{equation}
\left. \mu_k \right|_{max} \equiv \mu_k \simeq \dfrac{3}{8}\, \dfrac{\lambda}{\mu^4}\, g^2\, \zeta_3\mathcal{A}^2 \ \ .
\label{eq:instability_parameter_new}
\end{equation} 
This implies that the growing modes behave as 
\begin{displaymath}
 \delta\phi_k \ \sim \ e^{\frac{3}{4}\frac{\lambda}{\mu^4} g^3\,\zeta_3^{3/2}\, \mathcal{A}^2\, t} \ \ , \textrm{ \ \ which maximised yields } \  \delta\phi_k \ \lesssim \ e^{\frac{g \sqrt{\zeta_3}}{2}\, t} \ \ ;
\end{displaymath}
the maximum value will happen for the maximal speed of the brane moving along the throat. 
This results in particle production.

Comparing the two maximised instability parameters from the early (\ref{eq:instability_parameter_old}) and late (\ref{eq:instability_parameter_new}) time stages, we verify that $\mu_k > \mu_k^0 /2$ for some oscillation amplitude, $\mathcal{\bar{A}}$, of the inflaton field. This happens beyond a certain time. We conclude that the DBI correction boosts the explosive particle 
production when 
\begin{displaymath}
 \mathcal{A}<\mathcal{\bar{A}} \equiv \dfrac{1}{8}\, \sqrt{\zeta_3} \ = \ 0.125 \sqrt{\zeta_3} \ \ ,
\end{displaymath} 
ie, before the end of the oscillations, when $\mathcal{A}=\mathcal{A}_f\simeq0.025\sqrt{\zeta_3}$.

This behaviour should be true for adiabatic changes in $\mu$; otherwise, the growing mode would be of the form \[ \delta\phi_k \ \sim \ e^{\, \int{\mu_k(t) \, \omega dt}}.  \] However, we can compare the rate of the time evolution of $\mu$ with the typical frequency in our theory, $\omega$, as the suitable criteria for determining the adiabaticity of the resonance,
\begin{displaymath}
 \dfrac{\left|\dot{\mu}\right|}{\mu} \ll \omega \ \ ;
\label{eq:adiabaticity}
\end{displaymath} 
this is obeyed when the time evolution for the oscillation amplitude, $\mathcal{A}$, is such that
\begin{displaymath}
 \mathcal{A} \equiv \mathcal{A}(t) \ll \dfrac{1}{t^{1/2}} \ \ .
\end{displaymath} 
Provided this is true, then the resonant modes behave as described above. We expect $\mathcal{A}$ to be decreasing in time not only as a consequence of the Hubble friction term, but also due to the quartic interactions of $\chi$ which contribute to both rescattering and back-reaction. These factors all contribute to reduce the resonance effect.

\subsection{Is preheating efficiency undermined by the narrow resonance type?}
Since the resonance is narrow, rather than broad, we need to test whether or not it is efficient. Thus we need to compare the time each mode stays inside the resonance band. The necessary criteria is \cite{Kofman:1997yn}
\begin{displaymath}
 \tilde{q}^2 \, 2\omega \gtrsim H \ \ .
\end{displaymath}
This condition is verified for all amplitudes $\mathcal{A} \ll \mathcal{A}_c \sim 0.65\, \mu_{Pl} \sim 220\, \sqrt{\zeta_3}$, which is the case for expected values of $\sqrt{\zeta_3}$ \cite{Brandenberger:2008if}. Instead, if we used the Hubble parameter at the end of a slow-roll period \[H\simeq \dfrac{g}{\sqrt{3}} \, \dfrac{\zeta_3}{\mu_{Pl}}\ \ ,  \] then the criteria above would be satisfied for $\sqrt{\zeta_3}\ll \sqrt{3} \, \mu_{Pl}$, in agreement with the standard case.

\section{Reheating}
Due to the explosive behaviour of the resonance stage, the matter particles created via this process are far out of equilibrium. As previously discussed, not only matter particles but also inflaton particles are produced via the parametric resonance stage. Since this can have dangerous consequences, we need to provide a means for the inflaton particles to decay naturally into matter particles.
Thus we investigate possible decay channels to obtain standard reheating.
In this section, we also comment on the energy scales involved and how this 
affects the reheating temperature.

\subsection{Power spectrum constraints}
To proceed we first constrain the parameters of the theory, namely the symmetry breaking and cut-off energy scales normalised with respect to the 
CMBR \cite{Bailin:2004zd, Guth:1982ec}. For standard kinetic terms, the density perturbations are given by
\begin{equation}
 \dfrac{\delta\rho}{\rho}=-\dfrac{H^2}{\pi^{3/2}\, \dot{\phi}} \ \ , 
\end{equation} 
for a time evolving classical scalar field. For DBI models, due to the normalisation of the scalar field with respect to the string coupling, this becomes
\begin{equation}
 \dfrac{\delta\rho}{\rho}=-\dfrac{H^2}{\pi^{3/2}\, }\, \dfrac{\sqrt{g_s}}{\dot{\phi}} \ \ .
\end{equation} 
Since density perturbations are generated at the end of inflation and before 
preheating only the inflaton field will be considered. 
%Furthermore, as preheating takes place, the speed of the background field is monotically decreasing. 
In the regime for which $f\, \dot{\phi}^2\lesssim 2/3$, the maximal speed of the inflaton field is
\begin{displaymath}
\left| \dot{\phi} \right| \lesssim \left| \dot{\phi}_c \right| := \dfrac{1}{\sqrt{f}}\sqrt{\dfrac{2}{3}} \ \ .
\end{displaymath}
Since the inflaton field is rolling down the potential, we take the negative
root. At this stage the potential is
\begin{displaymath}
 V(\phi)\simeq\dfrac{g^2}{2}\, \zeta_3^2 \ \ ;
\end{displaymath} 
plugging in Friedmann's equation, we get
\begin{equation}
 H^2=\dfrac{1}{3\mu_{Pl}^2}\dfrac{V(\phi)}{g_s}=\dfrac{1}{6\mu_{Pl}^2}\dfrac{g^2\, \zeta_3^2}{g_s} \ \ .
\end{equation} 
Thus, matching with COBE data, it follows that
\begin{equation}
 \left(\dfrac{\delta\rho}{\rho} \right)^2=\dfrac{g^4}{24 \pi^3}\dfrac{\lambda}{g_s}\dfrac{\zeta_3^4}{\mu^4}\, \sim \, 4\times 10^{-10} \ \ .
\label{eq:power_spectrum}
\end{equation}
In \cite{Alishahiha:2004eh}, it was estimated that if the inflaton is to drive a purely de Sitter inflation, then this would lead to bounds on the 't Hooft coupling as $\lambda/g_s \sim 10^{12}$. Thus we expect, for reasonable values of $g$, $\zeta_3\ll \left(\mu\right) \, \mu_{Pl}$; this is consistent with our choice for the warp factor.

\subsection{Available decay channels}
Our aim is to propose a scenario in which decay channels are available as soon as the resonance stage is over so that the resonance process can be shut down at late times. On the other hand, these should not undermine it, which immediately discards the possibility of the inflaton decay via $\phi \phi \to \chi\chi$. Ultimately, as a natural consequence, we consider that any massive scalar matter particles produced here will eventually cascade down into the Standard Model particles and hence reproduce the content of our Universe. 

The proposed channels through which the residual energy density of the inflaton is 
transferred to matter are only efficient if
\begin{displaymath}
 \left< \chi^2\right> \ > \  \left< \phi^2\right> -\dfrac{\zeta_3}{2} +\dfrac{\tilde{\lambda}^2}{2g^2}\left<\phi\right>\  , \ \ \textrm{for the cubic channel}
\end{displaymath}
and
\begin{displaymath}
 \left< \chi^2\right> \ > \ \dfrac{h}{2g^2} \left< \phi\right>\ , \ \ \textrm{for the fermionic channel,}
\end{displaymath}
which is true when the fields reach their true vacuum values.

\subsubsection{Cubic interaction}\label{cubic}

We consider adding new interactions to the Lagrangian density $\mathcal{L}_0$ of the form $\tilde{\lambda}^2\phi \chi^2$, where $\tilde{\lambda}^2$ is the coupling constant, with dimensions of mass. It can be shown that the introduction of this term does not affect the preheating analysis provided
\begin{displaymath}
\dfrac{\sqrt{2}}{2}\, \dfrac{\tilde{\lambda}^2}{g^2} \ll \sqrt{\zeta_3}.
\end{displaymath}
We shall consider such regime. For this process, the decay rate is given by \cite{Kofman:1997yn}
\begin{equation}
 \Gamma\left(\phi \to \chi\chi \right)= \dfrac{\tilde{\lambda}^2}{8 \pi \omega} \ \ . 
\label{eq:decay_rate_cubic}
\end{equation} 
This should be compared with the resonance rate 
\begin{equation}
\dot{\mu} \sim \dfrac{\mu}{\triangle t} \sim \mu \omega \ \ .
\end{equation} 
Using (\ref{eq:instability_parameter_new}) we evaluate the condition $\Gamma \gtrsim \mu \omega$ for which the amplitude is
\begin{equation}
 \dfrac{\mathcal{A}^2}{g_s}< \dfrac{\tilde{\lambda}^2}{12\pi}\dfrac{\mu^4}{\lambda}\dfrac{1}{g^2\zeta_3}
\end{equation} 
For the final amplitude $\mathcal{A}_f=\sqrt{\zeta_3}/40$ (in $g_s$ units), the resonance turns off, and using (\ref{eq:power_spectrum}), we get the minimal value for the coupling constant for this decay channel:
\begin{equation}
 \tilde{\lambda}^2 > \left(7\times 10^{-9}\right)\, \dfrac{\mu_{Pl}^4}{g^2\, \zeta_3} \ \ .
\label{eq:coupling}
\end{equation} 
Both the estimates in (\ref{eq:power_spectrum}) and (\ref{eq:coupling}) will be relevant when evaluating the reheating temperature.

\subsubsection{Adding fermions}\label{fermions}
We could however expect other degrees of freedom to be present in additional decay channels. Thus we consider the coupling of the inflaton field to fermions. Although this is less efficient as far as the preheating mechanism goes, we can still evaluate the reheating temperature. We add the term $-h \, \bar{\psi}\, \psi \, \phi$ to the Lagrangian density, where $\psi$ denotes the fermion field and $h$ is a dimensionless coupling constant. The decay rate 
is given by
\begin{equation}
 \Gamma\left(\phi \to \bar{\psi}\psi \right)= \dfrac{h^2 \, \omega}{8 \pi } \ \ .
\label{eq:decay_rate_fermions}
\end{equation} 
If this is to prevail over the resonance stage at late times, then 
\begin{equation}
 \dfrac{\mathcal{A}^2}{g_s} \, < \, \dfrac{h^2}{3\pi}\, \dfrac{\mu^4}{\lambda} \dfrac{1}{g^2\, \zeta_3} \ \ .
\end{equation}

\subsection{Reheating temperature}
We can calculate the reheat temperature from the energy density. The 
energy density in particles is
\begin{displaymath}
 \rho \sim \dfrac{\Gamma_{eff}^2\, M_{Pl}^2}{24} \ \ ,
\end{displaymath}
where $\Gamma_{eff}$ is the effective decay rate. However, for relativistic 
particles in thermal equilibrium 
\begin{displaymath}
 \rho \sim \dfrac{\pi^2 }{30}\, N(T_R) \,T_R^4 \ \ ,
\end{displaymath}
where the number of degrees of freedom, $N(T_R)$, is of order $10^2$ or $10^3$
at the reheating temperature, $T_R$. Thus
\begin{equation}
 T_R \simeq 0.2 \, \sqrt{\Gamma_{eff}\, M_{Pl}} \ \ .
\end{equation} 

\subsubsection{Inflaton-Matter interaction}
Plugging in (\ref{eq:decay_rate_cubic}), then 
\begin{equation}
 T_R \simeq \, 2.1\times 10^{-2}\,\tilde{\lambda} \left(\dfrac{\mu_{Pl}}{g\sqrt{\zeta_3}}  \right)^{1/2} \ \ .
\label{eq:reheating_temperature}
\end{equation} 

From the COBE bound (\ref{eq:power_spectrum}), we get an expression for the symmetry breaking scale as a function of the coupling $g$ and the cut-off scale $\mu$:
\begin{equation}
 \zeta_3 \simeq 1.65 \times 10^{-5} \, \dfrac{\mu \, \mu_{Pl}}{g} \ \ .
\label{eq:zeta_condition}
\end{equation} 
Using (\ref{eq:coupling}), this gives the reheating temperature to be
\begin{equation}
 T_R \simeq 5\times 10^{-6} \,  \dfrac{\mu_{Pl}^{\ \ 5/2}}{\left(g^2\zeta_3  \right)^{3/4}} \ \ .
\end{equation} 
Inserting the constraint for the symmetry breaking scale in (\ref{eq:zeta_condition}), we finally get an estimate for the reheating temperature by eliminating the $\zeta_3$ dependence:
\begin{equation}
 T_R \simeq 2.1 \times 10^{-2} \, \dfrac{\mu_{Pl}^{\ \ 7/4}}{(g\, \mu)^{3/4}} \ \ .
\label{eq:reheating_temp_cubic}
\end{equation} 
So, for $g=8\times 10^{-1}$, then a reheating temperature of $T\sim 7\times 10^{16}GeV$ corresponds to a cut-off energy scale of $\mu \sim 9 \times 10^{-1} \mu_{Pl}$.
A general feature of (\ref{eq:reheating_temp_cubic}) is that for more warped throats the reheating temperature will be considerably larger. 
We note that for standard kinetic terms and coupling constant $g\sim 10^{-1}$, $T_R \sim 4.2 \times 10^{15} GeV$. Thus, DBI inflation seems to predict a
higher reheating temperature.

\subsubsection{Inflaton-Fermions interaction}
Proceeding in a similar fashion for the fermionic decay channel, taking 
$\lambda/g_s \sim 10^{12}$, we find that the reheating temperature is 
\begin{equation}
 T_R \sim 29.1 \, g^{3/2}\, \left( \dfrac{\mu_{Pl}}{\mu}  \right)^{5/4} \, \mu_{Pl} \ .
\end{equation} 
%for definiteness, for $g\sim 10^{-1}$, then $T_R\sim 1.9\times10^{19} GeV$ will correspond to a throat with cut-off energy scale $0.9\, M_{Pl}$. 
For $g \sim 10^{-3}$, one obtains $T_R \sim 6\times 10^{16} GeV$ for $\mu \sim 9\times 10^{-1}\, M_{Pl}$, which is again quite high.

%We cannot draw the comparison with the standard kinetic terms case here, as this breaks the Feynman perturbation expansion.

\section{Conclusions}

We have studied preheating and reheating in DBI inflation, coupling the inflaton to matter fields and using a hybrid potential. The expansion of the Universe was neglected in this work. By considering the DBI terms to be a small correction to the standard kinetic terms, we have determined the cosmological evolution of the inflaton as it oscillates around the minimum of the potential; we have also studied the evolution of the matter field. We have shown that preheating proceeds in the narrow resonance regime, which proves to be efficient in draining energy out of the inflaton field. Interestingly, DBI correction terms enhance the instability associated with the comoving modes of both fields after several oscillations. This suggests that non-standard kinetic terms favour efficient preheating by allowing stronger resonance of both the inflaton and the matter modes. 

In this paper, it was proposed that preheating was followed by the elementary theory of reheating. The conditions for this to happen constrained the coupling between the fields. Current bounds from the CMB spectrum were used to obtain estimates of the reheating temperature. This is a crucial step in order to achieve a perception of the energy scales in the theory. General estimates suggest that DBI inflation models give larger reheating temperature compared to theories with standard kinetic terms.

Although the work presented here is based on a toy model, it should give some insight into preheating in DBI inflation. We regard this as a first step. Future work should consider the expansion of the Universe and go beyond the approximation used in this paper. It would be interesting to draw a general conclusion of DBI inflation in (p)reheating with more general potentials \cite{common}.

\vspace*{1.5cm}

\acknowledgments
This work is supported in part by STFC.
R. H. R. acknowledges financial support from FCT (Funda\c{c}\~{a}o para a Ci\^{e}ncia e a Tecnologia - Portugal) through the grant SFRH/BD/35984/2007. We thank Philippe Brax, Carsten van de Bruck and in particular Joel Weller for discussions. We also thank Robert Brandenberger and Richard Easther for comments on an earlier version of this paper.

\appendix

\section{Classical Dynamics considerations}
We would like to obtain the asymptotic behaviour that $\phi \to 0$, at the bottom of the potential, using Classical dynamics principles. We intend to get such behaviour before $\phi$ reaches the true vacuum and before slow-roll conditions are no longer valid. Indeed, we start by considering the Lagrangian density in (\ref{eq:lagrangian}) and that both $\phi$ and $\chi$ will obey slow-roll conditions in this regime . In what follows and in conformity with the approximation scheme adopted, all quartic terms in the speed of $\phi$ and $\chi$ will be neglected.

The momenta conjugate to $\phi$ and $\chi$ are, respectively
\begin{displaymath}
 p_{\phi}:=\dfrac{\partial \mathcal{L}}{\partial \dot{\phi}} = \gamma \dot{\phi} \ \ \ \textrm{and} \ \ \  p_{\chi}:=\dfrac{\partial \mathcal{L}}{\partial \dot{\chi}} =  \dot{\chi} \ \ .
\end{displaymath}
Because the Lagrangian density is explicitly time-independent, it follows that the total energy is conserved 
\begin{equation}
 E=\dfrac{\gamma}{f}-\dfrac{1}{f}+\dfrac{\dot{\chi}^2}{2}+V\left(\phi\right) \ \ .
\label{eq:constant_energy}
\end{equation} 
Moreover, immediately before the beginning of the oscillations, we expect the time dependence of $\chi$ to be parametrised by that of $\phi$. Also, when the tachyonic instability sets in, the evolution of the fields will follow the linear trajectory (\ref{eq:trajectory}). In addition, writing for simplicity $V(\phi)=A+B\phi^2$, with $A=A(t)$ and $B=B(t)$, following the evolution of the matter field, then from (\ref{eq:constant_energy})
\begin{displaymath}
 \dfrac{1}{f\sqrt{1-f\dot{\phi}^2   }}=E+\dfrac{1}{f}-\left[\dot{\phi}^2 +A+B\phi^2  \right] \ .
\end{displaymath}
To proceed, we consider a ``short'' time evolution, $\triangle t$, so that $\dfrac{A}{\left|\dot{A}\right|} > \triangle t$ and $\dfrac{B}{\left|\dot{B}\right|} > \triangle t$. Redefining $T:= E+1/f-A \simeq constant$ and keeping the leading order terms
\begin{displaymath}
 \dot{\phi}^2 \simeq \dfrac{\left[T^2-\dfrac{1}{f^2}   \right]  -2B T\phi^2}{T\left(2+f T\right)} \ \ .
\end{displaymath}
Clearly, as we expect the inflaton field to be rolling down the potential, we take the negative root as the physical solution, so that
\begin{equation}
 \dot{\phi} \simeq -\dfrac{\left[T^2-\dfrac{1}{f^2}   \right]^{1/2}  \left[1- \dfrac{B T\phi^2}{T^2-\dfrac{1}{f^2}} \right]}{\left[T\left(2+f T\right)\right]^{1/2}} \ \ ,
\end{equation} 
for small $\phi$. At this stage, we require $T<0$ and $\left| f\, T  \right|>2$, so that
\begin{displaymath}
 \dot{\phi}=C\, \left( 1+D\phi^2  \right) \ \ , 
\end{displaymath}
with 
\begin{displaymath}
\left\{
\begin{array}{ll}
 \ \, C&= \sqrt{\dfrac{T^2-\dfrac{1}{f^2}}{T\left(2+f T\right)}} \\ \\
 \ \, D&= \dfrac{ B \, \left|T\right|}{T^2-\dfrac{1}{f^2}} \  \ .
\end{array}
\right. \ \ 
\end{displaymath}
This equation has solution
\begin{displaymath}
 \phi=-\dfrac{1}{\sqrt{D}}\tan{\left(C\, \sqrt{D} t+\beta\, \sqrt{D}  \right)} \ \ ,
\end{displaymath}
where $\beta$ is an integration constant which can be eliminated by requiring $\phi\left(t=t_f\right)=0$, for some time $t_f$; then
\begin{equation}
 \phi=-\dfrac{1}{\sqrt{D}}\, \tan{\left[C\, \sqrt{D}\, \left(t-t_f  \right)  \right]} \  \ .
\label{eq:tang_solution}
\end{equation} 
Manifestly, we are only able to study the evolution of $\phi$ in such a regime in less than half a period, ie, for $\triangle t <\dfrac{\pi}{2C\, \sqrt{D}}$. However, we can indeed check that $\phi \to 0$, given the choice of the initial condition above. This kind of behaviour was interestingly obtained \cite{Kecskemeti:2006cg} for the case of a finite throat, namely the Klebanov-Strassler throat, using the Hamilton-Jacobi method and neglecting any coupling of the inflaton with a matter field.

Importantly, one would like to connect this behaviour with the oscillatory one when the bottom of the potential is finally reached. For small angles $\tan{\theta}\simeq \sin{\theta}/\cos{\theta} \simeq\sin{\theta}$, which is, apart from a shift in the phase, the same oscillatory behaviour for the background solution. The requirements also agree with the ones made earlier to get (\ref{eq:tang_solution}) and certainly restrict the energy values for which this study applies.

\section{Insight on Hill's equation}
Given Hill's parameters in (\ref{eq:theta_values}), to evaluate the solutions of Hill's equation of the form
\begin{displaymath}
 \delta \phi_k(z) \ \sim \ e^{\mu_k \, z}\, \displaystyle{\sum_{r=-\infty}^{\infty}{c_{2r}\, e^{2r z i}}} \ + \ e^{-\mu_k \, z}\, \displaystyle{\sum_{r=-\infty}^{\infty}{c_{2r}\, e^{-2r z i}}} \ \ ,
\end{displaymath}
one usually applies Hill's method where recurrence formulae are arranged for the coefficients $c_{2r}$. 

We will now use the well understood behaviour of unstable solutions of Mathieu's equation in order to infer what happens with respect to Hill's equation. We start by noting that in the edges of the resonance bands $\mu_k=0$; it turns out that for the most important instability band and for $\theta_2 \ , \ \theta_4 \ < \ 1$
\begin{displaymath}
 \theta_0 \simeq 1 \pm \theta_2 \, \left[1\mp \dfrac{\left(\theta_4\right)^2}{6\, \theta_2} \right] \ \ ,
\end{displaymath} 
which, for $\theta_4 ^2 /\theta_2\ll 1$ (ie, well before the onset of reheating), essentially reduces to 
\begin{equation}
 \theta_0 \simeq 1 \pm \theta_2 \ \ \ .
\label{eq:theta_0}
\end{equation} 
From here, it also follows that, at lowest order, $\sqrt{\theta_0} \simeq 1\pm \theta_2/2$. Furthermore, from \cite{mclachlan:1965}, the instability parameter is given by
\begin{equation}
 \sin^{2}\left({i\mu_{k} \frac{\pi}{2}}\right)=\triangle(0) \, \sin^{2}\left({\sqrt{\theta_0} \, \frac{\pi}{2}}\right) \ \ ,
\label{eq:mu}
\end{equation}
where $\triangle(0)$ is the discriminant evaluated for $\mu=0$ (following from Hill's method); $\triangle (0)$ is exclusively determined by the parameters $\theta_0$, $\theta_2$ and $\theta_4$  and is approximately given by
\begin{displaymath}
 \triangle(0)\simeq 1+\dfrac{\pi}{4\, \sqrt{\theta_0}}\, \cot{\left(\sqrt{\theta_0}\, \dfrac{\pi}{2}  \right) \, \left[ \dfrac{\theta_2\, ^2}{1-\theta_0}  +  \dfrac{\theta_4\, ^2}{4-\theta_0}      \right]} \ \ ;
\end{displaymath} 
it can be shown that, approximately, this yields 
\begin{equation}
 \triangle(0) \simeq 1\mp \left(\dfrac{\pi}{2}\right)^2\, \left[\mp \left(\dfrac{\theta_2}{2}\right)^2 +\dfrac{\theta_2}{3}\, \left(\dfrac{\theta_4}{2}  \right)^2 +\dfrac{\left(\theta_2  \right)^3}{2}     \right]  \ \ .
\label{eq:triangle_0}
\end{equation} 
Using (\ref{eq:theta_0}) and (\ref{eq:triangle_0}), we find that from (\ref{eq:mu}), we may compute an approximate expression for the instability parameter for Hill's equation as 
\begin{equation}
 \mu_k \simeq \sqrt{\left(\dfrac{\theta_2}{2}\right)^2 -\left(\sqrt{\theta_0} -1 \right)^2  \mp \dfrac{\theta_2}{3}\, \left(\dfrac{\theta_4}{2}  \right)^2      \mp  \dfrac{\left(\theta_2  \right)^3}{2}     } \ \ .
\end{equation} 
This expression for $\mu_k$ is valid for $\theta_2, \theta_4 \ll 1$. Also, we may approximate it even further so as to recover the expression for the Floquet exponent in the case of Mathieu's equation:
\begin{equation}
 \mu_k \simeq \sqrt{\left(\dfrac{\theta_2}{2}\right)^2 -\left(\sqrt{\theta_0} -1 \right)^2          } \ \ .
\end{equation} 
This suggests that we evaluate the order of magnitude of $\theta_0$, $\theta_2$ and $\theta_4$ in the early and late time regimes separately and then assess their relative importance in order to make further approximations and eventually convert Hill's equation into the standard Mathieu's equation. This motivates the study on section \ref{efficiency}.

\bibliography{references}
\end{document}